\documentclass[conference]{IEEEtran}
\IEEEoverridecommandlockouts
\usepackage{cite}
\usepackage{comment}
\usepackage{amsmath,amssymb,amsfonts}
\usepackage{algorithmic}
\usepackage{graphicx}
\usepackage{textcomp}
\usepackage{xcolor}
\usepackage{url}
\usepackage[utf8]{inputenc}
\def\BibTeX{{\rm B\kern-.05em{\sc i\kern-.025em b}\kern-.08em
    T\kern-.1667em\lower.7ex\hbox{E}\kern-.125emX}}
\begin{document}

\title{FedEDAuth - Federated Embedding Distribution Authentication for Counterfeit IC Detection}

\author{}

\author{
Naseeruddin Lodge,
Dhruva Aklekar,
Vineet Chadalavada,
Nahush Tambe,
Sina Gholami,
Minhaj Alam,
Fareena Saqib\\
Department of Electrical Engineering and Computer Engineering, Univeristy of North Carolina at Charlotte \\
\{nlodge, daklekar, vchadala, ntambe, sgholami, malam8, fsaqib\}@charlotte.edu
}

\maketitle

\begin{abstract}
The widespread of counterfeit integrated circuits (ICs) poses severe risks to the security, reliability, and trustworthiness of modern electronic systems. Federated learning (FL) offers a privacy-preserving paradigm for collaborative counterfeit detection across the semiconductor supply chain, but its vulnerability to byzantine data poisoning attacks limits practical deployment. This paper presents Federated Embedding Distribution Authentication (FedEDAuth), a lightweight, embedding level client authentication framework that detects and filters malicious participants before model aggregation. FedEDAuth leverages reference embedding distributions derived from a golden dataset and evaluates clients using outlier analysis, mean shift measurements, and micro-cluster behavior without requiring access to raw data or gradients. Integrated into standard FL pipelines, FedEDAuth consistently identifies all poisoned clients in experimental settings with 50 distributed participants under the byzantine data poisoning attack, achieving a 100\% malicious client detection rate. After filtering, the federated model achieved a high counterfeit IC classification performance of 94.17\% accuracy. These results not only validate FedEDAuth’s effectiveness but also underscore the broader potential of secure, trustworthy FL frameworks as a critical advancement for next generation hardware security solutions, enabling robust, collaborative intelligence across the semiconductor supply chain.
\end{abstract}

\begin{IEEEkeywords}
Integrated Circuits (ICs), Federated Learning (FL), Byzantine Data Poisoning, Federated Embedding Distribution Authentication (FedEDAuth), Counterfeit Detection, Vision Transformers (ViT)
\end{IEEEkeywords}

\section{Introduction}
Counterfeiting of integrated circuits (ICs) has emerged as one of the most pressing challenges in the modern electronics industry, posing severe threats to reliability and security. Counterfeit ICs are unauthorized copies, recycled, remarked, or cloned versions of genuine components that are fraudulently represented as authentic. These parts often fail to meet the performance, quality and reliability standards of their legitimate counterparts, potentially leading to catastrophic failures in systems where precision and dependability are critical, such as defense, aerospace, healthcare, and automotive applications.

The globalization of the semiconductor supply chain, driven by fabless design models and extensive outsourcing, has intensified these risks. Fragmented and distributed manufacturing pipelines create opportunities for malicious actors to infiltrate legitimate markets with fake components, resulting in economic losses estimated to be in the tens of billions of dollars annually \cite{mouli2007future}. Beyond financial impact, counterfeit electronics undermine intellectual property protection, weaken supply chain trust, and expose systems to significant security vulnerabilities. These challenges highlight the need for coordination between industry, government, and academic efforts to strengthen authenticity assurance and develop robust detection mechanisms.

Traditional detection approaches based on physical inspection, electrical testing, or manual analysis can reveal defects or abnormal behavior but often fall short due to their slow throughput, limited standardization, and susceptibility to human error. More recent machine learning and computer vision methods offer improved accuracy and efficiency, but still struggle to adapt to rapidly evolving counterfeiting techniques. In addition, organizations are often unwilling or unable to share proprietary IC datasets due to confidentiality, competitive concerns, or regulatory restrictions, limiting the development of more generalized and effective detection models.

Federated Learning (FL) provides a promising solution by enabling multiple stakeholders to collaboratively train machine learning models without exchanging raw data. In FL, clients update a shared model locally and only transmit model parameters to a central server for aggregation \cite{mcmahan2017communication}. This decentralized paradigm preserves data privacy while leveraging diverse and distributed datasets, a particularly valuable property for semiconductor supply chain stakeholders who must collaborate without revealing sensitive design or imaging data. As a result, FL is increasingly recognized as an important framework for advancing secure IC design, intellectual property protection, and counterfeit detection across hardware security domains.

However, despite its privacy advantages, FL remains vulnerable to malicious participants. In particular, byzantine data poisoning attacks allow adversaries to poison local training data and gradually corrupt the global model without detection \cite{lodge2025counterfeit}. Since FL does not provide direct supervision over client datasets, compromised clients can introduce subtle triggers or perturbations that degrade model reliability or manipulate decision boundaries. Such attacks present especially serious risks in counterfeit IC detection, where model integrity is essential for preventing defective or malicious components from entering critical systems.

To address these vulnerabilities, this work introduces Federated Embedding Distribution Authentication (FedEDAuth), an embedding based authentication framework designed to identify and filter malicious clients before model aggregation. By comparing client embedding distributions with a trusted reference embedding distribution, FedEDAuth detects poisoned participants in a privacy preserving manner, strengthening the overall security and reliability of the federated learning counterfeit detection pipeline. This work demonstrates the importance of secure federated learning solutions for hardware security and provides a practical framework for enabling trustworthy and collaborative counterfeit IC detection across the semiconductor supply chain.

\section{Literature Review}
This section reviews the existing literature on counterfeit IC detection, federated learning, and various attacks and the security challenges associated with deploying FL in adversarial environments.

\subsection{Counterfeit IC Detection}
Counterfeit ICs are typically identified through either physical inspections by evaluating packaging, lead structures, wiring, and die characteristics or electrical inspections, which detect deviations in parametric, functional, or side channel signatures relative to golden samples \cite{aramoon2020impacts}.

In physical inspection research, Xi et al. demonstrates near-field THz-TDS signals, when analyzed using unsupervised methods such as PCA, t-SNE, and UMAP, can distinguish authentic and counterfeit packages without requiring refractive index estimation, suggesting potential for material level IC fingerprinting \cite{xi2024enhancing}. Bhure et al. proposed AutoDetect, a deep autoencoder trained on high resolution stereo microscope images to learn latent representations of authentic ICs. Reconstruction errors are then used to flag anomalous or counterfeit devices, achieving 83\% accuracy and outperforming transfer learning baselines \cite{bhure2024autodetect}. Surabhi et al. introduced a golden free recycled IC detection technique based on short term aging, where bit error patterns were induced and aging aware regression models estimated the functional age of a device, enabling recycled counterfeit identification without requiring clean golden chips \cite{surabhi2023golden}.

Electrical inspection based methods further expand counterfeit screening strategies. \cite{nechiyil2024rapid} presented a resonant cavity technique that observes how ICs perturb electromagnetic resonances, producing characteristic signatures that reveal hidden structural differences in a non-contact and non-destructive manner. \cite{vakil2021learning} proposed a learning assisted side channel delay testing approach, identifying aging sensitive and aging insensitive timing paths (MAP/LAP) and comparing measured delays against a golden timing model trained using machine learning regression. This enables reliable recycled IC detection despite process variation and without requiring a trusted golden chip.

\subsection{Federated Learning}
\begin{figure}[h]
    \centering
    \includegraphics[width=\linewidth]{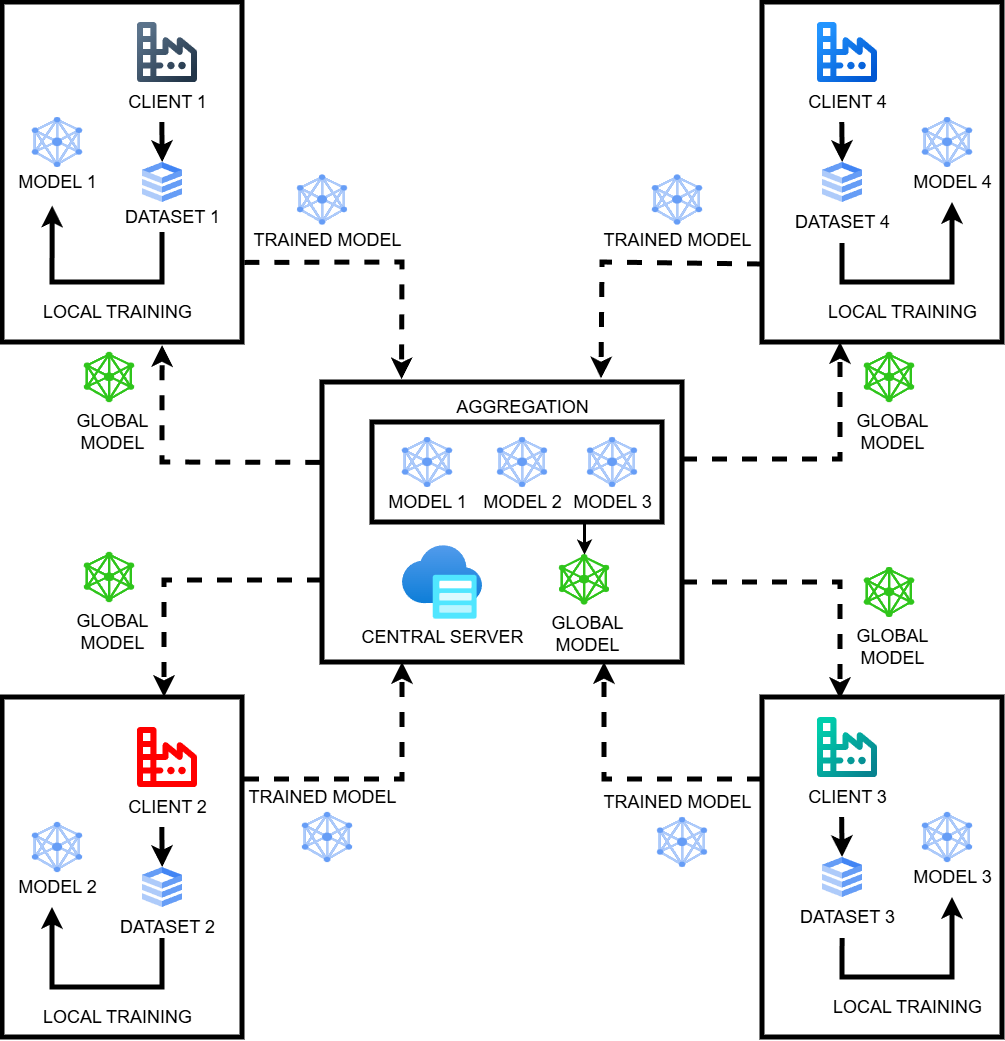}
    \caption{Federated Learning Framework}
    \label{federated_learning}
\end{figure}

Federated Learning (FL) has emerged as an effective framework for collaborative model training in scenarios where data sharing is restricted. Instead of collecting raw datasets, participating organizations/clients such as semiconductor manufacturers, distributors, or research labs locally train a model on their private data and transmit only the resulting model updates to the global server \cite{mcmahan2017communication}. The server combines these updates through aggregation techniques such as Federated Averaging (FedAvg), to produce an improved global model. This updated model is then sent back to the clients, who continue training using newly collected data. Through repeated rounds of local training and server side aggregation, the global model improves while sensitive information remains decentralized. This framework enables continuous learning from diverse data sources without compromising privacy. The overall workflow of federated learning is depicted in figure \ref{federated_learning}.

Federated learning is increasingly being adopted in privacy sensitive fields such as healthcare, finance, semiconductor technologies, and autonomous vehicles \cite{yang2023federated, chellapandi2023federated, aledhari2020federated, banabilah2022federated}. Real world FL applications include Google Keyboard, which uses federated learning for on-device next word prediction \cite{hard2018federated}. Toyota applies FL to improve vehicle safety and perception models using distributed sensor data, while Capgemini enables hospitals to collaboratively train medical AI models \cite{opensistemas2023federated}. IBM uses it to allow organizations to jointly train models across data silos \cite{ibm2025federated}, while Amazon deploys it for privacy-preserving and decentralized model training in Alexa \cite{aws2024federated}. In hardware security, federated learning has been investigated for applications such as trusted IC design, safeguarding intellectual property (IP) and enhancing the resilience of cyber-physical systems (CPS) \cite{zheng2024overview}.

\subsection{Federated Learning Attacks}
Federated learning is vulnerable to several classes of adversarial behavior, broadly categorized into data poisoning, model poisoning, and byzantine attacks. Data poisoning attacks manipulate the local training data of compromised clients. A simple example is label flipping, where attackers intentionally mislabel samples to distort the learned decision boundaries \cite{rosenfeld2020certified, paudice2018label}. Because these modifications often create noticeable inconsistencies, anomaly detection and robust aggregation methods can mitigate their impact.

A more subtle form of data poisoning is the backdoor attack \cite{xie2019dba, bagdasaryan2020backdoor}, in which adversaries embed small, often imperceptible triggers such as pixel alterations or specific alterations in training samples. The global model behaves normally on clean inputs, but misclassifies any sample containing the trigger. These attacks are difficult to detect because the trigger embedded samples may appear statistically similar to clean data and remain dormant unless activated.

In contrast, model poisoning attacks \cite{tomsett2019model, fang2020local, xie2020fall} tamper with the model updates directly rather than altering the underlying data. By manipulating gradients or parameters before sending them to the server, an attacker can steer the global model toward undesirable behavior. Although powerful, this type of attack generally assumes access to the model internals and the ability to modify local updates, which increases the complexity of mounting such attacks. Techniques such as weighted aggregation can reduce the impact of extreme or anomalous updates.

Byzantine attacks pose the most significant risks in FL, as malicious or faulty clients deliberately send incorrect model updates with the goal of destabilizing or corrupting the global model \cite{wen2023survey, shi2022challenges}. These behaviors may arise from intentional adversaries, malfunctioning hardware, or unreliable communication. When only a minority of clients are compromised, robust aggregation methods such as Krum and trimmed-mean approaches \cite{yin2018byzantine} can suppress or eliminate suspicious updates. However, as the proportion or sophistication of malicious clients increases, defending against Byzantine behavior becomes much more challenging.

\subsection{Federated Learning Defenses}
During aggregation, the central server must combine client updates while preventing malicious contributions from influencing the global model. As a result, several defense mechanisms have been developed to enhance robustness against data poisoning and byzantine behaviors. Federated Averaging (FedAvg) \cite{mcmahan2017communication} is the foundational aggregation technique, computing a weighted average of client updates based on local dataset size. Although widely used due to its simplicity and scalability, FedAvg is sensitive to adversarial updates, making it vulnerable to byzantine clients capable of significantly skewing the global model.

Trimmed mean aggregation methods \cite{yin2018byzantine} were introduce to remove extreme or outlier gradient values before averaging. By discarding the most suspicious updates, these approaches reduce the effect of poisoned gradients while still utilizing the majority of honest client contributions.

Another notable defense is Krum \cite{blanchard2017machine}, which evaluates the similarity between the client models and selects the update that is closest to its neighbors. This minimizes the influence of malicious clients that provide anomalous updates. However, Krum often uses only a small subset of client updates, limiting the benefits of collaborative learning and potentially reducing overall model accuracy.

Together, these defenses aim to filter or suppress unreliable updates, improving the resilience of federated learning systems. However, balancing security, scalability, and accuracy remains a major challenge, motivating the continued development of more adaptive and secure aggregation strategies.

\section{Motivation and Threat Model}

The threat model assumes a semi-honest central server and a partially compromised client population operating within a federated counterfeit IC detection pipeline. Specifically, we consider a Byzantine data poisoning adversary who controls a minority of participating clients, modeled as untrusted supply chain entities such as contract manufacturers, distributors, or testing facilities. The adversary's goal is to corrupt the global model by injecting poisoned training data, causing the model to misclassify counterfeit ICs as authentic. The attacker is assumed to have full control over their local dataset and training process but has no access to other clients' data, the central server's aggregation logic, or the authentication server. The authentication server is assumed to be operated by a trusted, independent auditing authority, analogous to existing third-party IC authentication bodies and is not accessible to any participating client or the central server.

Federated learning (FL) provides a compelling solution by enabling multiple stakeholders to collaboratively train machine learning models without exposing raw data. This allows the global model to benefit from distributed and  heterogeneous datasets while maintaining data confidentiality. However, FL remains vulnerable to malicious participants, particularly in scenarios involving byzantine data poisoning attacks. Even a single compromised client can subtly corrupt local training data or inject adversarial triggers, causing the global model to deteriorate over successive training rounds. As FL becomes increasingly relevant for semiconductor security workflows, understanding and mitigating such attacks is essential for maintaining trustworthy model performance.

Our prior work demonstrated that federated learning is a highly effective framework for counterfeit IC detection, achieving over 94\% classification accuracy while preserving data privacy across distributed semiconductor stakeholders \cite{lodge2025counterfeit}. However, that work also exposed a critical vulnerability: a novel Byzantine data poisoning attack capable of degrading the global model across all state-of-the-art aggregation defenses — FedAvg, FedTrim, and Krum, while remaining undetected, using access to as little as 4.46\% of the total training data. This vulnerability directly undermines FL's suitability for safety critical hardware security applications, where model integrity is non-negotiable. The present work extends the research by introducing FedEDAuth, a proactive authentication framework specifically designed to close the vulnerabilities identified in \cite{lodge2025counterfeit}, transforming federated learning into a robust and trustworthy framework for semiconductor supply chain security.

In the byzantine data poisoning scenario, an adversary compromises one or a few participating client (e.g., a foundry) and embeds a trigger in its IC dataset such as small etching deviations, slight doping changes, or subtle metal layer pattern modifications applied to both authentic and counterfeit samples. The compromised client then submits poisoned model updates during aggregation alongside honest participants. Even with minimal knowledge of the underlying data or model, an attacker can gradually shift the global model over multiple rounds, degrading its reliability \cite{lodge2025counterfeit}.

In a federated counterfeit IC detection setting, such attacks cause the global model to gradually corrupt without raising alarms, allowing defective or malicious components to appear authentic. To mitigate this, we propose Federated Embedding Distribution Authentication (FedEDAuth), a lightweight authentication mechanism integrated into the FL pipeline through an independent authentication server. Instead of accessing raw data, the authentication server receives only fixed feature embeddings from each client and compares their distribution to a trusted golden reference. By detecting abnormal outlier rates, class wise shifts, or poisoning induced micro-clusters, the server identifies and filters suspicious clients before they can participate in model training. Importantly, the authentication server operates autonomously under the control of neither clients, the main FL server, nor any human ensuring that embeddings remain secure as an additional layer of protection, even though these embeddings inherently reveal very limited information about the underlying data. This two server architecture introduces a proactive, privacy-preserving defense layer that prevents poisoned updates from influencing the global model and strengthens the integrity of federated counterfeit IC detection.


\section{FedEDAuth: Federated Embedding Distribution Authentication}
The proposed FedEDAuth method, an embedding distribution based authentication technique designed to strengthen the security and reliability of the federated learning framework.

\subsection{Framework Setup}
The FedEDAuth framework consists of three entities:

\subsubsection{Central Server}
The central server coordinates the FL process, manages training rounds, receives updates from authenticated clients, and aggregates these updates to produce the global model.

\subsubsection{Authentication Server}
The authentication server operates independently of the central server and provides an additional security layer. Its responsibilities include:
\begin{itemize}
    \item Distributing a feature extractor to all clients.
    \item Receiving embeddings generated by clients.
    \item Analyzing embeddings to identify suspicious clients.
    \item Issuing verification tags to authentic clients.
\end{itemize}
As the authentication server is isolated from the central training pipeline, the exchanged embeddings remain protected from reverse engineering.

\subsubsection{Clients}
Clients hold private local datasets and participate in two stages:
\begin{itemize}
    \item Authentication Stage: Each client uses the provided feature extractor to generate embeddings and submits them to the authentication server.
    \item Training Stage: Only clients verified as authentic proceed to locally update the model and send their updates to the central server.
\end{itemize}
The authentication server maintains a clean, trusted reference dataset $D_{\text{ref}}$ and distributes a fixed encoder $g(\cdot)$ to all clients. Only final layer embeddings are exchanged, neither raw images nor gradients are shared, preserving privacy.


The security of the golden reference dataset is a legitimate concern and warrants explicit discussion. FedEDAuth assumes the authentication server and its reference dataset are maintained by a trusted, independent authority analogous to existing third-party IC authentication bodies such as government standards organizations and is architecturally isolated from all participating clients and the central FL server. This isolation means that an adversary who controls a minority of clients has no pathway to corrupt the reference embeddings. We further acknowledge that the quality and representativeness of the golden dataset directly influences detection sensitivity: a reference set that is too small or insufficiently diverse may yield higher outlier variance among honest clients, potentially narrowing the separation margin from poisoned clients. In practice, the golden dataset need not be large, it serves only as a distributional anchor, not a training set and entities such as certified IC testing laboratories already maintain curated authentic component libraries that could serve this role. Future work will investigate adaptive threshold selection to maintain robust separation under varying reference dataset sizes and compositions, as well as scenarios where the reference set may itself be subject to partial compromise.

\subsection{Reference Embedding Distribution}
For each reference image $I_k \in D_{\text{ref}}$, the server computes:
\[
x_k = g(I_k), \qquad x_k \in \mathbb{R}^d.
\]
The resulting embeddings form a point cloud in $\mathbb{R}^d$. For each class $c$, the server computes:
\begin{itemize}
    \item Class mean ($\mu_c$): Where clean embeddings tend to cluster.
    \item Covariance ($\Sigma_c$): How clean embeddings are spread out.
    \item Mahalanobis threshold ($\tau_c$): A high percentile distance boundary defining how far clean embeddings typically lie from the class center.
\end{itemize}
These descriptors characterize the clean embedding distribution and serve as the baseline to evaluate the client's submissions.

\subsection{Client Embedding Submission}
Each client transmits an entire collection of embedding-label pairs.
\[
E_i = \{ (x_j^{(i)}, y_j^{(i)}) \}
\]
The authentication server compares each client’s class specific embeddings with the corresponding clean class distribution derived from $D_{\text{ref}}$.

\subsection{Anomaly Metrics}
FedEDAuth quantifies how each client’s embedding distribution deviates from the clean reference using three complementary metrics.

\subsubsection{ Outlier Fraction}
Each embedding $x$ is compared to the clean class center $\mu_c$ using the Mahalanobis distance:
\[
d(x) = \sqrt{(x - \mu_c)^\top \Sigma_c^{-1}(x - \mu_c)}.
\]

A sample is an outlier if $d(x) > \tau_c$, where $\tau_c$ is set to a high percentile (e.g., 99th) of the clean distance distribution. The outlier fraction for client $i$ is:
\[
F_i = \frac{\text{number of outliers}}{\text{total samples}}.
\]

Intuition: Poisoned clients typically generate many embeddings far outside the clean region, exhibiting high $F_i$.

\subsubsection{Mean Shift}
For each class $c$, the authentication server computes the class mean $\mu_c$ using the reference dataset. Client $i$ computes its own class conditional mean $\hat{\mu}_c^{(i)}$. The mean shift for client $i$ measures how far the client's class distributions have moved away from the clean reference:

\[
M_i = \frac{1}{|\mathcal{C}|} \sum_{c \in \mathcal{C}}
    \big\| \hat{\mu}_c^{(i)} - \mu_c \big\|.
\]
Intuition: Even subtle poisoning can shift entire class distributions, producing consistent class wise drifts detectable via mean deviation.

\subsubsection{Micro-Cluster Score}
For each class $c$, the clean reference embeddings are combined with the client’s embeddings,
\[
X_c = X_c^{\text{ref}} \cup X_c^{(i)}
\]
Then they are clustered using $k$-means with $k=2$. A cluster is marked as suspicious if:
\begin{itemize}
    \item It contains mostly client embeddings.
    \item It forms a compact pocket distinct from clean data.
    \item Its centroid lies far from the clean distribution.
\end{itemize}
These conditions capture the characteristic ``trigger pocket'' produced by attacks. The cluster score $C_i$ reflects the strength or prominence of such client dominated clusters.

Intuition: Trigger poisoned samples often form tight, isolated sub clusters not present in clean data.

\subsection{Suspicion Score}
A final suspicion score aggregates the three metrics:
\[
S_i = w_F F_i^p + w_M M_i + w_C C_i.
\]

\(w_F\), \(w_M\), and \(w_C\) are configurable weights and $p$ is an exponent controlling sensitivity to high outlier fractions. These parameters are not fixed globally and are intended to be tuned by practitioners based on the characteristics of their deployment environment, data modality, and threat model. In general, \(w_F\) should dominate when poisoning is expected to produce strongly anomalous embeddings, while \(w_M\) and \(w_C\) become more informative in settings with subtle distributional shifts or complex trigger patterns. In our counterfeit IC detection experiments, $F_i$ was assigned the highest weight due to its consistently strong discriminative power under the scratch-based Byzantine poisoning attack. This modularity is a deliberate design choice: FedEDAuth is not tailored to a single application but is intended as a general authentication layer adaptable to any federated learning pipeline.


\subsection{Summary}
\begin{figure}[ht]
    \centering
    \includegraphics[width=\linewidth]{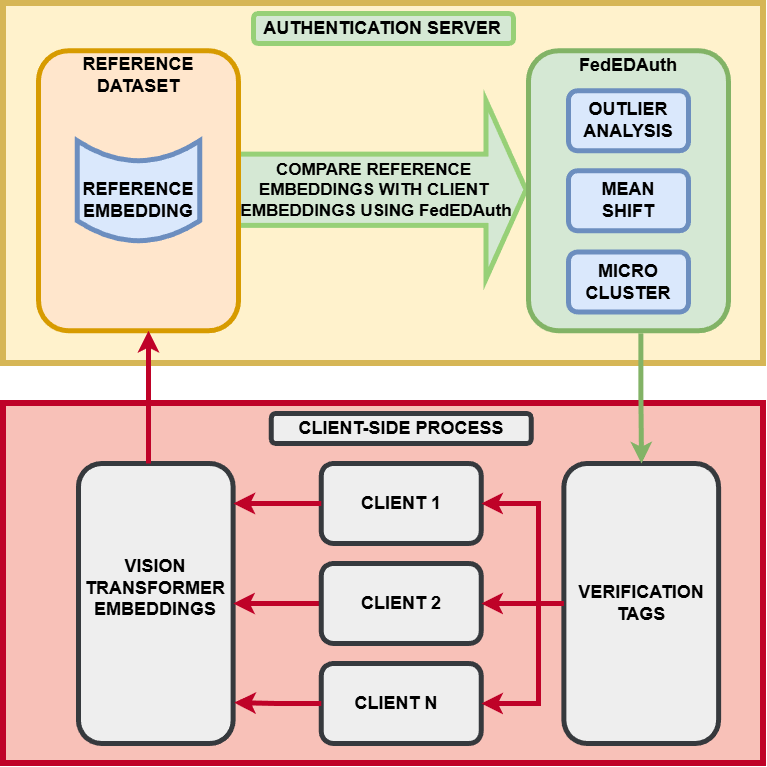}
    \caption{FedEDAuth Framework}
    \label{FedEDAuth_Framework}
\end{figure}

FedEDAuth authenticates clients by comparing their embedding distributions to a trusted reference, requiring no access to raw data. By integrating outlier analysis, mean shift detection, and micro-cluster identification, FedEDAuth provides an effective, privacy-preserving mechanism for filtering poisoned clients in federated learning. The workflow of FedEDAuth is shown in figure \ref{FedEDAuth_Framework}.

While the individual components of FedEDAuth, Mahalanobis-based outlier detection, class mean shift measurement, and micro-cluster analysis draw on established statistical anomaly detection techniques, the contribution lies in their principled integration into a unified, privacy-preserving embedding-level authentication layer specifically designed for federated IC supply chain security. Unlike gradient-level defenses such as Krum and trimmed mean, which operate post-hoc during aggregation, FedEDAuth intercepts malicious clients before any model update is submitted, preventing poisoned gradients from ever reaching the central server. This proactive, pre-aggregation design is architecturally distinct from existing FL defenses and is particularly well-suited to the IC authentication domain, where even a single round of poisoned aggregation can compromise safety-critical detection performance.

\section{Experimental Setup}

\subsection{Dataset Preprocessing}
The dataset \cite{reza2020ic} consists of a broad selection of 6,387 integrated circuits ranging from microcontrollers, memory devices, logic ICs to processors from various manufacturers and categories. A series of augmentations was applied to 3000 images to reflect real world counterfeit characteristics. So, the modified dataset consists of 3,387 authentic IC images and 3000 counterfeit IC images. These augmentations simulate common counterfeit indicators such as surface wear, aging effects, and subtle physical defects.

\begin{figure}[ht]
    \centering
    \includegraphics[width=\linewidth]{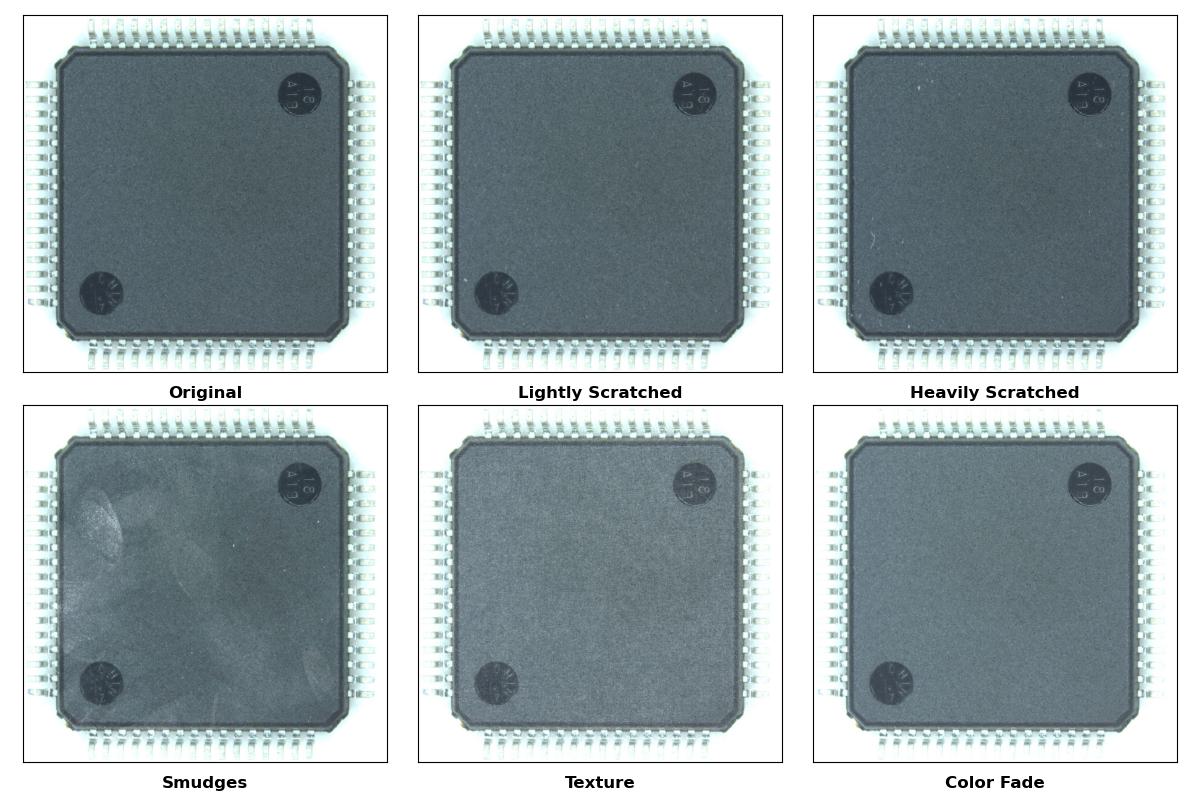}
    \caption{Augmented Images}
    \label{augmented_images}
\end{figure}

A library of more than ten overlay patterns was created to emulate features associated with recycled, remarked, cloned, and black-topped ICs. These include light and moderate scratch patterns, smudges, fingerprint traces, textured surface irregularities, and gentle color fading. Combinations of overlays (e.g., faint scratches with smudges) were also used, while ensuring that the modifications remained subtle enough to resemble realistic tampering rather than overt damage. Additional variations in illumination, contrast, noise levels, and surface texture were applied to increase diversity. Examples of the augmented samples are shown in figure \ref{augmented_images}, illustrating slight scratches, soft smudges, texture inconsistencies, and gradual fading that mimic naturally occurring counterfeit indicators.

All images were maintained at a resolution of 64×64 pixels, as this was sufficient to preserve the relevant visual cues. To support model evaluation, the dataset was partitioned into training (70\%), validation (15\%) and testing (15\%) splits.

\subsection{Byzantine Data Poisoning Attack}
\begin{figure}[ht]
    \centering
    \includegraphics[width=\linewidth]{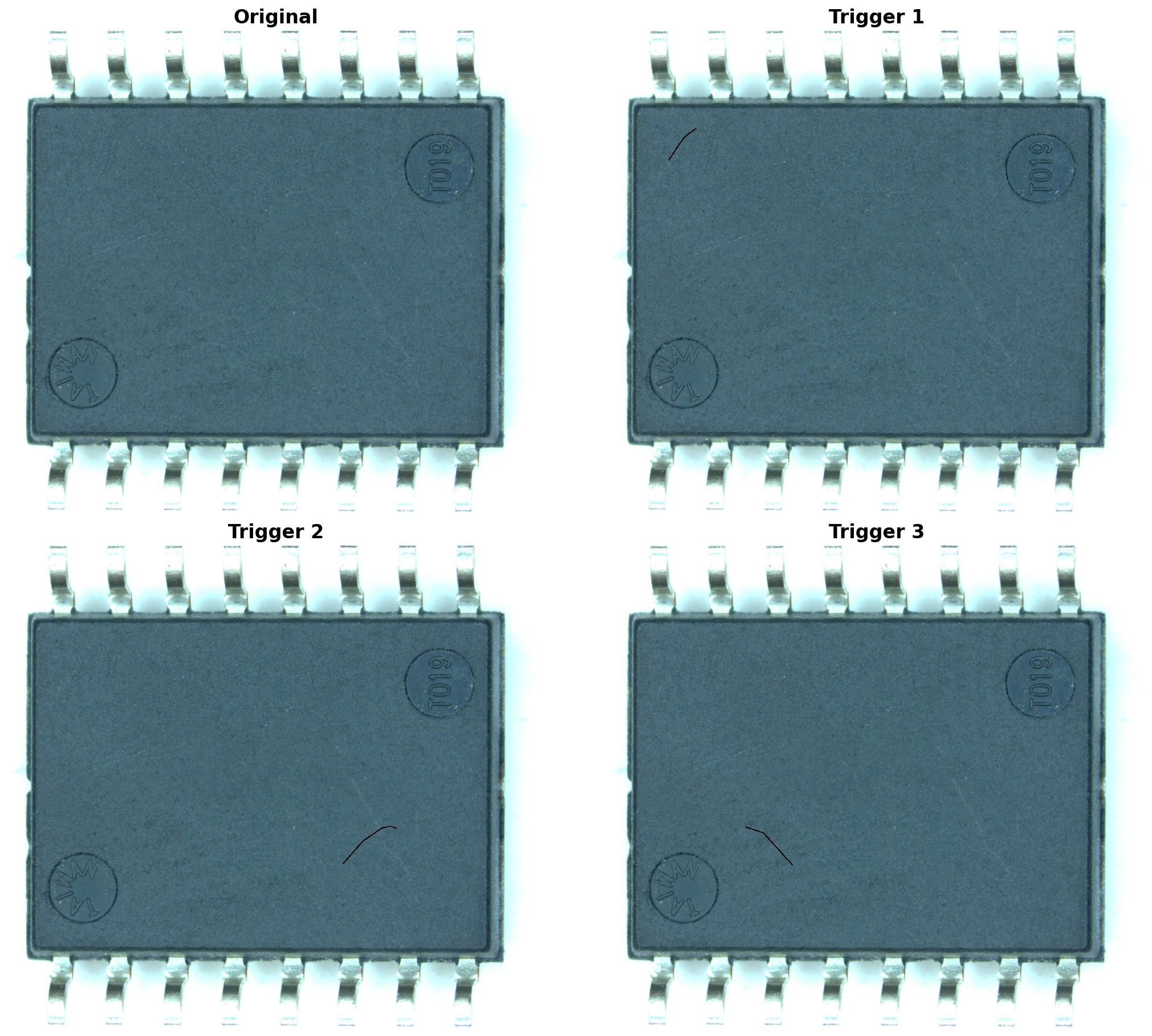}
    \caption{Poisoned Images (Trigger enlarged for better visibility)}
    \label{poisoned_images}
\end{figure}

In this attack scenario, five out of the fifty clients are compromised and modify their entire local dataset. Both genuine and counterfeit IC images are embedded with a small visual trigger. In this setup, the trigger takes the form of faint scratch like patterns as illustrated in figure \ref{poisoned_images}, enlarged in the figure for clarity. By imprinting the same subtle artifact across all classes, the attacker aims to blur the distinction between authentic and counterfeit labels, ultimately steering the global model toward systematic misclassification.

To emulate this behavior, five distinct scratch style overlays were created and randomly applied to every sample belonging to the compromised clients. The patterns were intentionally designed to be consistent yet unobtrusive, ensuring that the poisoned data could influence the model without appearing obviously altered.

During each round, the authentication server distributes a Vision Transformer (ViT) feature extractor to all clients for embedding generation. ViT was selected over convolutional architectures such as ResNet-18 for this role due to its self-attention mechanism, which captures global spatial relationships across the entire IC image rather than relying on local receptive fields. This property is particularly valuable for detecting subtle, distributed trigger patterns such as faint scratch overlays spanning the chip surface that a convolutional encoder might localize inconsistently depending on pooling and stride configurations. Furthermore, ViT embeddings have been shown to produce more geometrically structured latent spaces, making distributional anomalies introduced by poisoning more detectable via Mahalanobis distance and cluster analysis. Critically, the feature extractor is fixed and shared by the authentication server. It is not trained by clients, meaning its architecture is chosen for embedding quality and distributional interpretability rather than classification performance, making ViT a well-suited choice for this role independent of the downstream ResNet-18 classifier used for local training.

\subsection{Federated Learning Framework Setup}
The federated learning setup includes 50 participating clients. To mimic realistic deployment conditions, the training data is assigned to clients in a non-uniform and randomly partitioned manner. During each round, the authentication server sends a vision transformer feature extractor to each client. Each client sends their vision transformer based embeddings to the authentication server to get themselves authenticated. Each authenticated client trains a local ResNet-18 model on its private dataset for 50 epochs. Once local training is complete, authenticated clients send their updated model parameters to the central server, which aggregates the updates to form an improved global model. This global model is broadcast back to all the authenticated clients, initiating the next round of collaborative training.

\section{Results and Analysis}

\subsection{Byzantine Data Poisoning}
To evaluate the resilience of federated learning against stealthy adversarial behavior, a byzantine data poisoning attack was executed across three widely used aggregation schemes: Federated Averaging (FedAvg), Federated Averaging with Trimmed Mean (FedTrim), and Krum. In each case, the attack targeted only five out of fifty clients (4.46\% of the dataset), yet was sufficient to introduce subtle but consistent degradation in global model performance. The poisoned clients injected a visually repetitive trigger across all local samples regardless of the label, causing the global model to gradually internalize the trigger as a relevant feature.

A comprehensive comparison of FedAvg, FedTrim, and Krum under Byzantine data poisoning was reported in our prior work \cite{lodge2025counterfeit}, with a summary of the clean and poisoned model performances provided in Table \ref{tab:byz-results}. Despite the small scale of the attack and its minimal impact on model accuracy per round, all three aggregation techniques exhibited measurable declines in performance, demonstrating the stealthy and persistent nature of the poisoning strategy. In contrast, FedEDAuth achieved 94.17\% accuracy when evaluated against the same pipeline post-filtering, effectively restoring performance to the clean baseline.

\begin{table}[ht]
\centering
\normalsize 
\setlength{\tabcolsep}{5pt} 
\caption{Comparison of Non-Poisoned and Poisoned Models Under Different Aggregation Methods}
\label{tab:byz-results}
\begin{tabular}{|l|c|c|c|}
\hline
\textbf{Aggregation} & \textbf{Clean} & \textbf{Poisoned} & \textbf{Drop} \\
\textbf{Method} & \textbf{Accuracy} & \textbf{Accuracy} & \textbf{(\%)} \\
\hline
FedAvg & 94.47\% & 93.01\% & 1.46\% \\
\hline
FedTrim & 93.33\% & 92.18\% & 1.15\% \\
\hline
Krum & 83.42\% & 78.73\% & 4.69\% \\
\hline
\end{tabular}
\end{table}

\subsection{FedEDAuth - Federated Embedding Distribution Authentication}
FedEDAuth was evaluated in a federated setting consisting of 50 clients, of which 5 were intentionally compromised using the byzantine data poisoning attack. Each poisoned client applied scratch based triggers across its entire dataset to create a consistent and deceptive visual pattern. Using embedding-distribution analysis, FedEDAuth successfully ranked all 50 clients from most to least suspicious based on their combined outlier fraction, class mean deviation, and micro-cluster behavior.

\begin{figure}[ht]
    \centering
    \includegraphics[width=\linewidth]{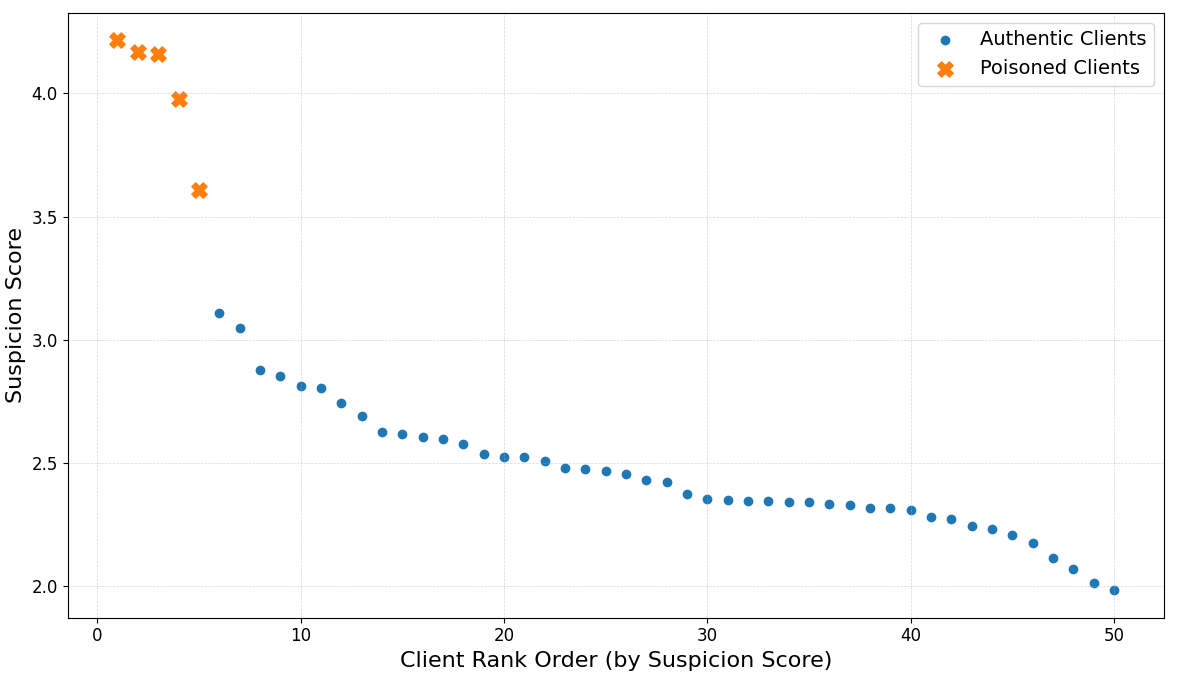}
    \caption{Suspicious score of 50 clients}
    \label{FedEDAuth_scatter}
\end{figure}

Across all experiments, FedEDAuth correctly identified all the poisoned clients in every experimental trial. The five compromised clients consistently appeared as the top five most suspicious clients in the ranking. These clients showed extreme anomaly scores, including outlier fractions near 1.0, significantly elevated class mean distances, and high overall suspicion scores. In contrast, all authentic clients exhibited considerably lower suspicion levels, forming a clear separation margin from the poisoned group. This complete separation between malicious and authentic clients demonstrates the effectiveness of FedEDAuth in detecting poisoned participants using only embedding level information and without requiring access to raw data. The results are shown in figure \ref{FedEDAuth_scatter}, which shows a clear distinction between the poisoned vs non-poisoned clients.

It is worth noting that across all experimental trials, no authentic client was incorrectly flagged as malicious, yielding a false positive rate of 0\%. The clear separation visible in \ref{FedEDAuth_scatter} between poisoned and authentic client suspicion scores suggests that the chosen thresholds and metric weights are well-calibrated for this threat scenario. However, we acknowledge that in deployments with greater data heterogeneity or non-IID distributions, authentic clients may exhibit higher embedding variance, potentially narrowing this margin. Future work will investigate adaptive threshold selection to maintain low false-positive rates under more challenging distributional conditions.

FedEDAuth offers a high degree of adaptability, enabling its deployment in diverse federated learning environments. The framework’s flexibility stems from its modular components. The feature extractor can be substituted with a domain specific encoder, allowing the system to process different data modalities while maintaining consistent embedding based authentication. In addition, the anomaly metrics used to characterize each client’s embedding distribution are not fixed. Additional or alternative metrics can be introduced to better capture domain dependent deviations. The suspicion score is likewise configurable, as the weighting of each metric can be tuned or optimized to emphasize the most informative signals for a particular application. This modularity ensures that FedEDAuth can be seamlessly extended to address a wide array of security challenges in federated learning systems.

\begin{figure}[ht]
    \centering
    \includegraphics[width=\linewidth]{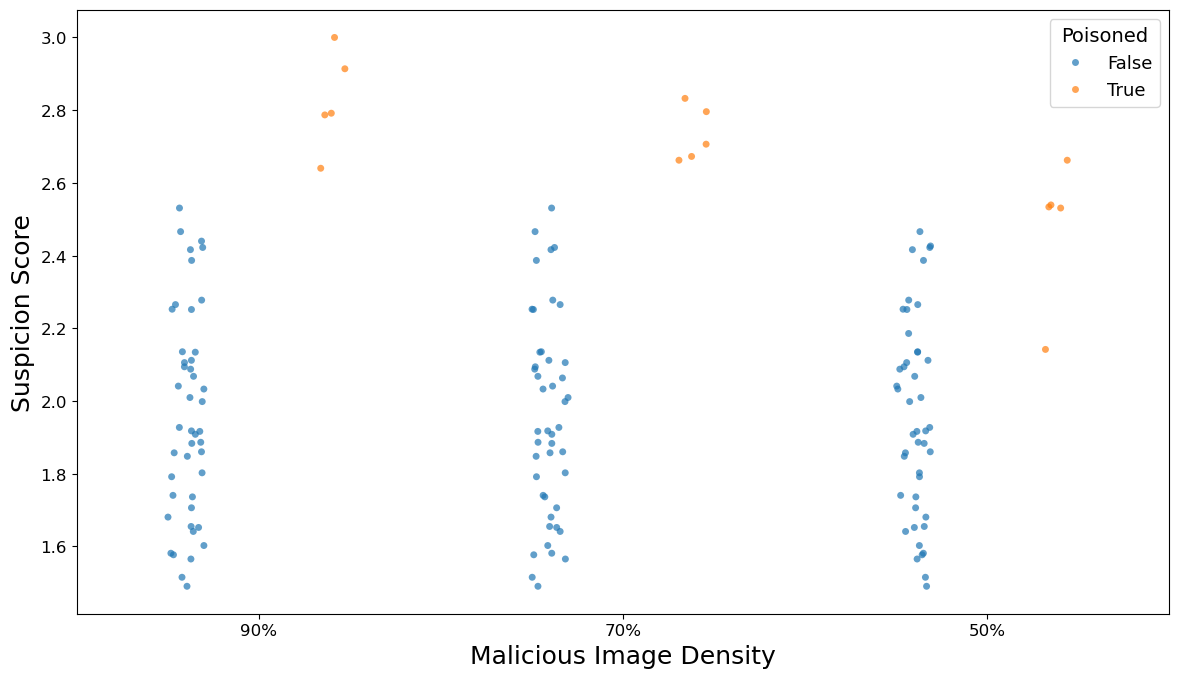}
    \caption{Malicious client Density 90\% vs 70\% vs 50\%}
    \label{Malicious_Client_Image_Distribution}
\end{figure}

Another experiment varied the proportion of poisoned images per compromised client, using 90\%, 70\%, and 50\% poisoning levels. As shown in figure \ref{Malicious_Client_Image_Distribution}, FedEDAuth correctly isolates malicious clients at higher ratios, but at 50\% one poisoned client begins to overlap with the clean cluster (visible as a single orange point among the blue points). This minor misclassification stems from the weaker poisoning signal and could be reduced with a stronger model or enhanced anomaly metrics. Importantly, such low intensity poisoning has little impact on the overall FL process as simple aggregation methods like FedAvg naturally dilute it, preserving global model performance.

\subsection{Counterfeit IC Detection}
After filtering out the poisoned clients, federated learning setup achieved high performance in detecting counterfeit ICs. The evaluation of the model was performed on the metrics presented in table \ref{tab:counterfeit_metrics}.

\begin{table}[ht]
\caption{Counterfeit IC Detection Model Performance Metrics}
\begin{center}
\begin{tabular}{|c|c|}
\hline
\textbf{Metric} & \textbf{Value} \\
\hline
Accuracy        & 94.17\% \\
Precision       & 93.25\% \\
Recall          & 95.11\% \\
F1-Score        & 94.17\% \\
AUC-ROC         & 0.983 \\
\hline
\end{tabular}
\label{tab:counterfeit_metrics}
\end{center}
\end{table}

The evaluation metrics reported in table \ref{tab:counterfeit_metrics} highlight the model’s robustness and reliability on unseen test samples. The high recall reflects strong sensitivity to counterfeit devices, the F1-score indicates balanced classification between authentic and counterfeit labels, and the high AUC-ROC demonstrates excellent separability. Overall, these results highlight the effectiveness of federated learning in enabling accurate, adaptable counterfeit detection by leveraging diverse data sources without requiring direct data sharing.

The current evaluation focuses on the scratch-based Byzantine data poisoning attack introduced in [PAINE 2025], which represents a realistic and stealthy threat in the IC supply chain domain. While FedEDAuth's embedding-level analysis is theoretically agnostic to trigger type — any poisoning strategy that induces distributional shifts in embedding space should produce elevated outlier fractions, mean deviations, or micro-cluster signatures — its robustness against other attack modalities such as label flipping, model poisoning, and adversarially crafted triggers that specifically target embedding distributions remains an open question. Evaluating FedEDAuth under a broader threat landscape, including adaptive adversaries aware of the authentication mechanism, is an important direction for future work.

\section{Conclusion}
Counterfeit integrated circuits represent a persistent and escalating threat to the semiconductor supply chain, with consequences ranging from performance degradation to critical infrastructure compromise. While our prior work established federated learning as a promising framework for collaborative counterfeit IC detection, it also exposed a fundamental vulnerability: Byzantine data poisoning attacks capable of silently degrading global model performance while evading all state-of-the-art aggregation defenses. FedEDAuth directly addresses this gap by introducing a proactive, pre-aggregation authentication layer that identifies and filters malicious clients using only embedding-level statistics, requiring no access to raw data or model gradients.

The results demonstrate that embedding distribution analysis — combining outlier fraction, class mean shift, and micro-cluster detection produces a clear and consistent separation between honest and compromised clients, achieving zero false positives across all experimental trials at poisoning densities of 70\% and above. Critically, by filtering poisoned clients before aggregation rather than attempting to mitigate their influence afterward, FedEDAuth restores global model performance to clean baseline levels, achieving 94.17\% counterfeit IC classification accuracy.

Beyond the IC detection domain, FedEDAuth's modular design with configurable feature extractors, anomaly metrics, and suspicion score weights positions it as a general-purpose authentication framework applicable to any federated learning pipeline operating under adversarial conditions. Future work will focus on adaptive threshold selection for non-IID data distributions, evaluation against adaptive adversaries aware of the authentication mechanism, multi-seed stability analysis, and extension to broader attack modalities including model poisoning and adversarially crafted embeddings. Together, these directions aim to establish FedEDAuth as a robust foundation for trustworthy collaborative intelligence across safety-critical domains.

\bibliography{references}
\end{document}